\documentclass[prl,aps,amsmath,amssymb,twocolumn,superscriptaddress,longbibliography]{revtex4-1}
\usepackage{graphicx,bm,color,appendix}
\usepackage[colorlinks,bookmarks=true,citecolor=blue,linkcolor=blue,urlcolor=blue]{hyperref}
\usepackage{times}
\usepackage{amsmath,amssymb}
\usepackage{comment}
\usepackage{appendix}

\graphicspath{{./figs/}}

\usepackage{xcolor}
\definecolor{orange}{rgb}{1,0.5,0}

\begin{document}

\title{Conformal four-point correlators of the 3D Ising transition via the quantum fuzzy sphere}

\author{Chao Han}
\affiliation{Westlake Institute of Advanced Study,	Westlake University, Hangzhou 310024, China }

\author{Liangdong Hu}
\affiliation{School of Sciences,	 Westlake University, Hangzhou 310024, China }

\author{W. Zhu}
\email{zhuwei@westlake.edu.cn}
\affiliation{School of Sciences,	 Westlake University, Hangzhou 310024, China }
\affiliation{Westlake Institute of Advanced Study, Westlake University, Hangzhou 310024, China }

\author{Yin-Chen He}
\email{yhe@perimeterinstitute.ca}
\affiliation{Perimeter Institute for Theoretical Physics, Waterloo, Ontario N2L 2Y5, Canada}

\begin{abstract}
In conformal field theory (CFT), the four-point correlator is a fundamental object that encodes CFT properties, constrains CFT structures, and connects to the gravitational scattering amplitude in holography theory. However, the four-point correlator of CFTs in dimensions higher than 2D remains largely unexplored due to the lack of non-perturbative tools. In this paper, we introduce a new approach for directly computing four-point correlators of 3D CFTs. Our method employs the recently proposed fuzzy (non-commutative) sphere regularization, and we apply it to the paradigmatic 3D Ising CFT. Specifically, we have computed three different four-point correlators: $\langle \sigma\sigma\sigma\sigma\rangle$, $\langle \sigma\sigma\epsilon\epsilon\rangle$, and $\langle \sigma\sigma T_{\mu\nu} T_{\rho\eta}\rangle$. Additionally, we verify the crossing symmetry of $\langle \sigma\sigma\sigma\sigma\rangle$, which is a notable property arising from conformal symmetry. Remarkably, the computed four-point correlators exhibit continuous crossing ratios, showcasing the continuum nature of the fuzzy sphere regularization scheme. This characteristic renders them highly suitable for future theoretical applications, enabling further advancements and insights in 3D CFT.
\end{abstract}

\maketitle

Conformal field theories (CFTs) are a fascinating class of quantum field theories with an elegant mathematical structure and wide-ranging applications, from classical and quantum phase transitions in condensed matter physics~\cite{Sachdev_book,Cardy_book} to quantum gravity~\cite{Maldacena_AdSCFT}. However, CFTs beyond 2D ~\cite{Belavin1984} pose theoretical challenges due to their strongly interacting nature. Developing non-perturbative numerical or analytical tools to study CFTs in 3D and higher dimensions continues to be an ongoing challenge for various physics communities.

The power and elegance of conformal symmetry are revealed through the highly constrained correlators of primary operators in CFTs \cite{yellowbook}. Specifically, the two-point (scalar) correlators are given by $\langle \phi_i(\bm x_1) \phi_j(\bm x_2)\rangle = \frac{\delta_{ij}}{x_{12}^{2\Delta_i}}$, while the three-point (scalar) correlators are expressed as $\langle \phi_i(\bm x_1) \phi_j(\bm x_2) \phi_k(\bm x_3)\rangle = \frac{f_{ijk}}{x_{12}^{\Delta_i+\Delta_j-\Delta_k} x_{23}^{\Delta_j+\Delta_k-\Delta_i} x_{31}^{\Delta_k+\Delta_i-\Delta_j}}$~\cite{polyakov1970conformal}. These correlators are completely determined up to universal data, such as scaling dimensions $\Delta_i$ and operator product expansion coefficients $f_{ijk}$.
In contrast, the four-point correlator will not be completely fixed. However, by virtue of conformal symmetry, it can be reduced to a universal function of two variables that depends on the theory and operators involved, e.g., for scalars,
\begin{equation}\label{eq:4pt_definition}
\langle {\phi}(\bm x_4) {\phi}(\bm x_3) {\phi}(\bm x_2) {\phi}(\bm x_1) \rangle = \frac{g(u, v)}{x_{12}^{2\Delta} x_{34}^{2\Delta} },
\end{equation}
where $u=\frac{x_{12}^2 x_{34}^2}{x_{13}^2 x_{24}^2}$ and $v=\frac{x_{14}^2 x_{23}^2}{x_{13}^2 x_{24}^2}$ are called crossing ratios.
Importantly, the four-point correlator satisfies crossing symmetry, which is obtained by exchanging points in Eq.\eqref{eq:4pt_definition}. For example, $\bm x_1 \leftrightarrow \bm x_3$ gives $u^{-\Delta} g(u, v) = v^{-\Delta} g(v, u)$. This crossing equation, also known as the bootstrap equation\cite{polyakov1974nonhamiltonian}, imposes highly stringent constraints on the conformal data of CFTs.
Solving these equations constitutes the conformal bootstrap program~\cite{polyakov1974nonhamiltonian,RMP_CB}. In general, there is an infinite number of bootstrap equations due to the vast number of global conformal primaries, making the analytical implementation of the conformal bootstrap program a daunting task. However, the conformal bootstrap program has seen successful applications in 2D CFTs, thanks to the presence of a larger emergent local conformal symmetry~\cite{Belavin1984}.
Excitingly, recent progress in 3D has demonstrated the ability to obtain strong numerical constraints from a limited number of $2\sim3$ bootstrap equations. This advancement has notably led to the determination of world-record precise critical exponents for the 3D Ising CFT \cite{RMP_CB}.

The four-point correlator plays a central role in the realm of CFTs. It not only constrains conformal data but also encodes these data, which can be extracted using the inversion formula~\cite{Simon2017Inversion}. Moreover, in the context of the AdS/CFT correspondence~\cite{Maldacena_AdSCFT}, the four-point correlator of a CFT is related to the scattering amplitude of particles in the dual quantum gravitational theory in AdS spacetime~\cite{Polchinski1999Smatrix,Susskind1999Holography}.
However, little is known about the four-point correlator of 3D CFTs due to the lack of computational methods. For the paradigmatic 3D Ising CFT, the only non-perturbative computation of the four-point correlator was achieved through an indirect reconstruction using the conformal block expansion based on numerical conformal bootstrap data \cite{Rychkov2017_4pt}.
In this Letter, we present a new approach by utilizing the recently proposed \emph{fuzzy (non-commutative) sphere regularization}~\cite{ZHHHH2022}, enabling us to directly compute four-point correlators of the 3D Ising CFT. Specifically, we compute $\langle\sigma\sigma\sigma\sigma\rangle$, $\langle \sigma \sigma \epsilon\epsilon\rangle$, and $\langle \sigma\sigma T_{\mu\nu} T_{\rho\eta}\rangle$, with the last one being beyond the state-of-the-art bootstrap computation. Furthermore, we directly verify the crossing symmetry of $\langle \sigma\sigma\sigma\sigma\rangle$ in our computation.
Notably, unlike the traditional lattice regularization, the fuzzy sphere regularization preserves the continuum nature even for finite physical volumes. As a result, the four-point correlator computed using the fuzzy sphere approach is a continuous function of crossing ratios, making it amenable to various applications, including the application of the inversion formula.

\emph{Fuzzy sphere regularization.--}The fuzzy sphere regularization scheme~\cite{ZHHHH2022} involves studying a continuum strongly interacting quantum mechanical model, where particles (e.g., fermions) reside on the surface of a two-sphere $S^2$ in the presence of a magnetic monopole. Specifically, for the 3D Ising CFT, one can study a  $2+1$D transverse Ising model on the sphere with a charge $4\pi s$ monopole at the origin.
Besides the kinetic energy of fermions, the system is characterized by a continuous Hamiltonian: 
\begin{align} \label{eq:Hamcontinuum}
 \int d\Omega_a d\Omega_b \,  U(\Omega_{ab}) 2n^\uparrow(\Omega_a)n^\downarrow(\Omega_b) - h \int d \Omega \, n^x(\Omega). 
\end{align}
The model consists of spin operators, namely $n^{\uparrow(\downarrow)} = \psi^\dag_{\uparrow(\downarrow)} \psi_{\uparrow(\downarrow)}$, and $n^\alpha(\Omega) = \bm{\psi}^\dag(\Omega) \sigma^a \bm{\psi}(\Omega)$, where $\bm{\psi}^\dag(\Omega) = (\psi_\uparrow^\dag(\Omega), \psi_\downarrow^\dag(\Omega))$ represents the spinful (non-relativistic) electron operator, and $\sigma^{x,y,z}$ are the Pauli matrices.
The first term, $2n^\uparrow(\Omega_a)n^\downarrow(\Omega_b) = n^0(\Omega_a)n^0(\Omega_b) - n^z(\Omega_a)n^z(\Omega_b)$, describes the Ising-type density-density interaction between electrons located at spatial points $\Omega_a = (\theta_a, \varphi_a)$ and $\Omega_b = (\theta_b, \varphi_b)$ on a sphere with a radius $R$.
The interaction term $U(\Omega_{ab}) = \frac{g_0}{R^2}\delta(\Omega_{ab}) + \frac{g_1}{R^4}\nabla^2\delta(\Omega_{ab})$ is taken to be a local and short-ranged interaction, ensuring that the phase transition is described by a local theory. The second term represents a transverse field.
At the half-filling, the transverse field $h$ triggers a phase transition from a quantum Hall ferromagnet \cite{Girvin2000} with spontaneous $\mathbb Z_2$ symmetry breaking to a quantum paramagnet, which falls into the 2+1D Ising universality class. 

The emergent conformal symmetry of the transition has been convincingly demonstrated through radial quantization. Both the scaling dimensions~\cite{ZHHHH2022} and the operator product expansion coefficients~\cite{hu2023operator} of conformal primary operators have been computed with high accuracy. For a detailed description of the fuzzy sphere regularization, we refer to Ref.~\cite{ZHHHH2022}.
In this paper, we will use the same interaction form $U(\Omega_{ab})$ as presented in Ref.~\cite{ZHHHH2022,hu2023operator}. Additionally, we will reexamine the critical field strength $h_c$ by employing a new quantity, namely the dimensionless two-point correlator, which will be described in detail below.

\begin{figure}[t]
\includegraphics[width=0.45\linewidth]{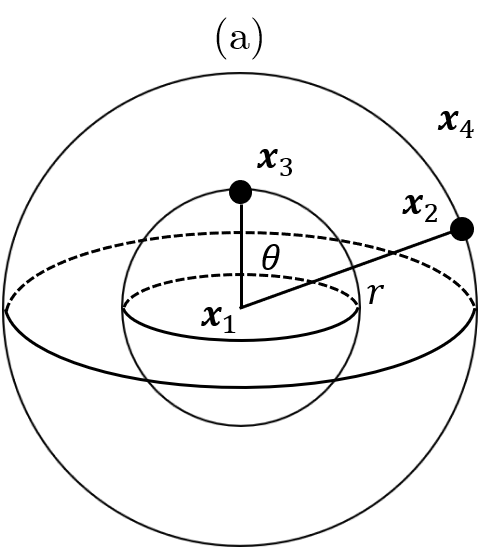}
\includegraphics[width=0.49\linewidth]{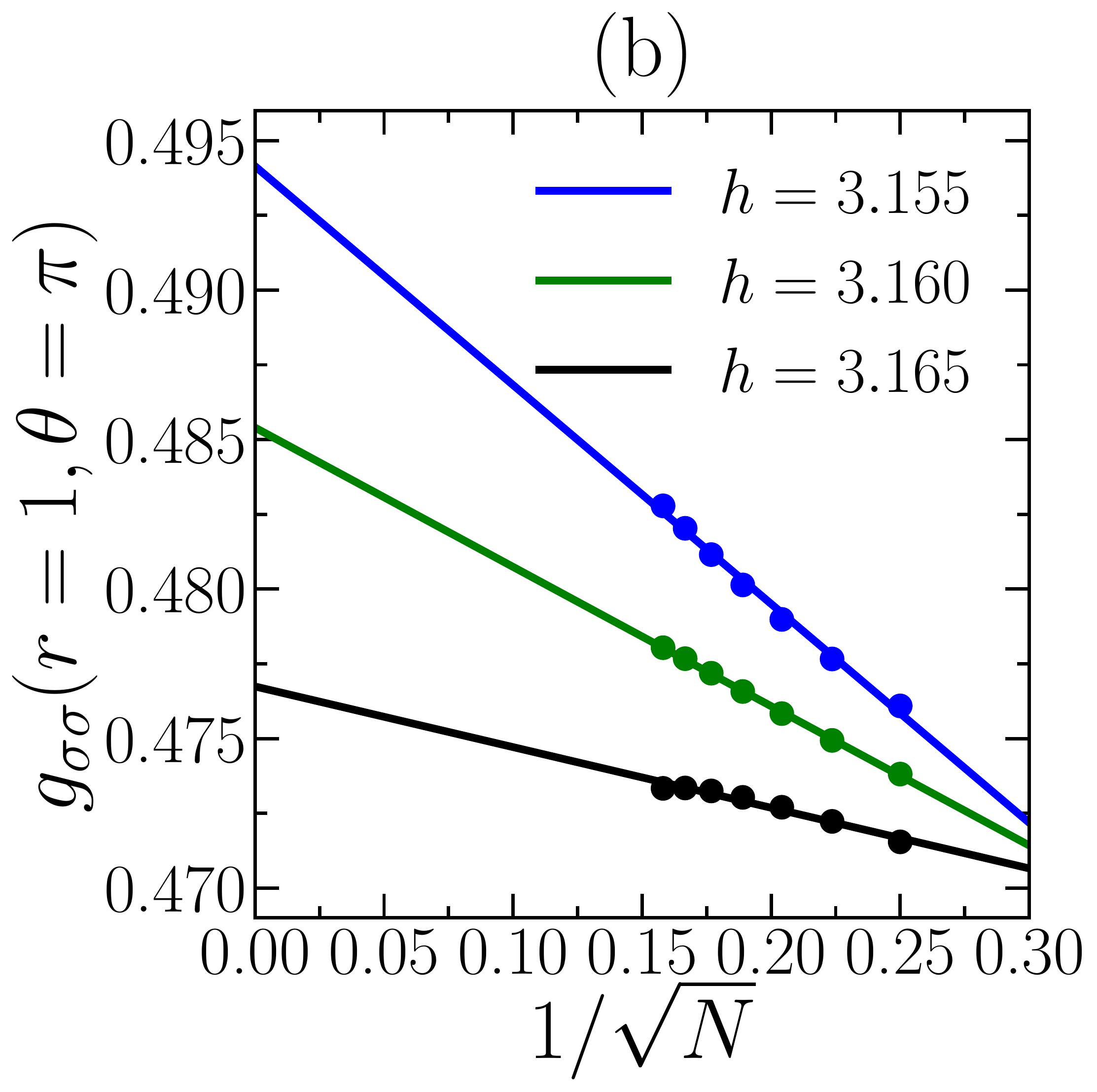}
\includegraphics[width=0.49\linewidth]{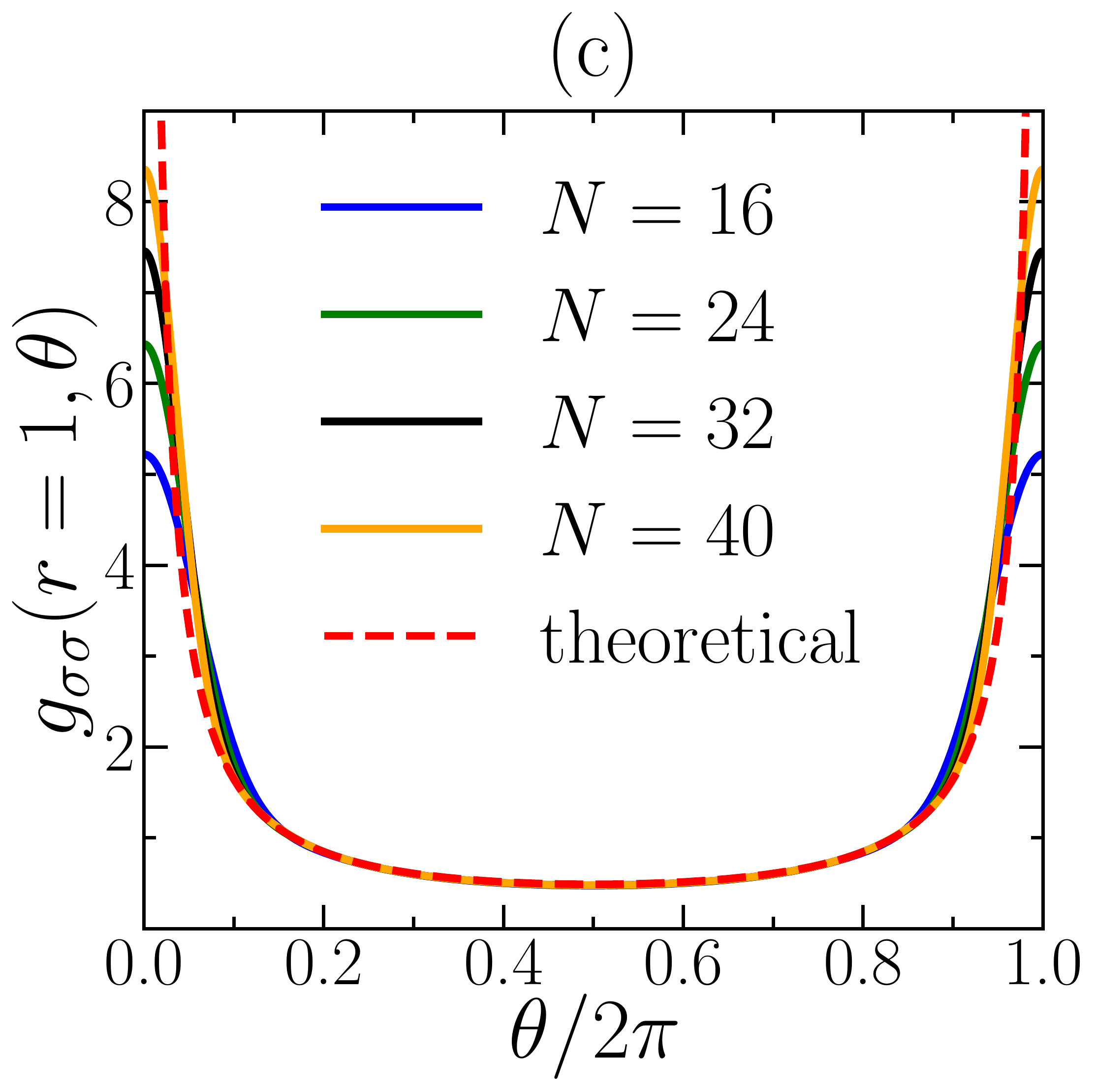}
\includegraphics[width=0.49\linewidth]{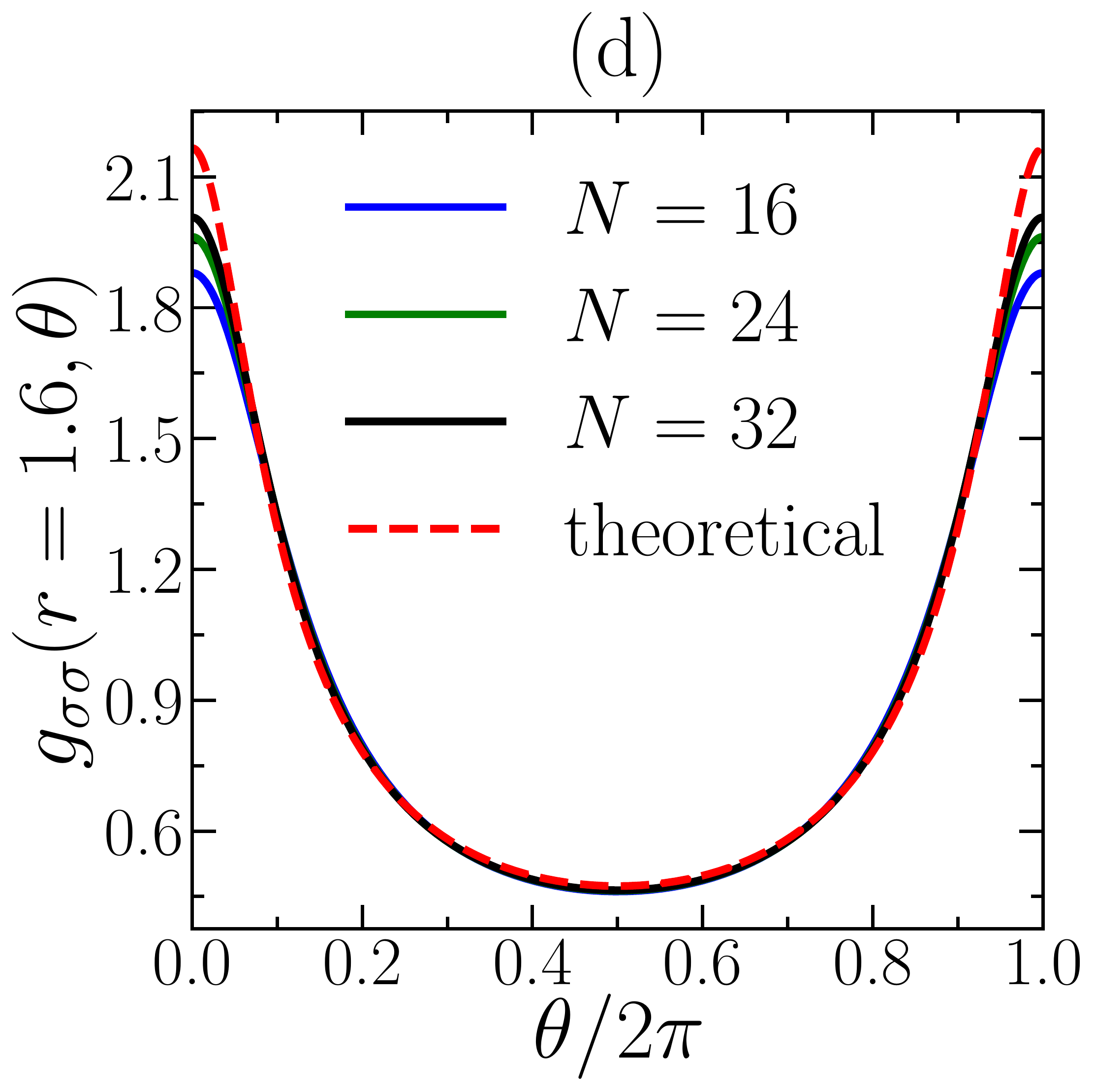}
\caption{(a) In spherical coordinates, $\bm x_1$, $\bm x_2$, $\bm x_3$, and $\bm x_4$ are respectively positioned as follows: $\bm x_1$ is located at the origin, $\bm x_2$ is set at the point $(r, \theta, \varphi=0)$, $\bm x_3$ is positioned at the north pole of the unit sphere $(r=1, \theta=0, \varphi=0)$, and $\bm x_4$ is placed at infinity.
(b) The finite-size scaling of the two-point correlator $g_{\sigma\sigma}(r=1,\theta=\pi)$ is shown for different values of $h$, namely $h=3.155, 3.16, 3.165$. The theoretical value at the critical point is approximately 0.487577, which is close to the value obtained at $h=3.16$.
(c) The angle dependence of the two-point correlator $g_{\sigma\sigma}(r=1,\theta)$ is plotted for system sizes $N=16-40$, while (d) shows $g_{\sigma\sigma}(r=1.6,\theta)$ for system sizes $N=16-32$. The red dashed line represents the theoretical prediction. The discrepancy is more pronounced for small values of $\theta$ due to the singularity at $\theta=0$ in the thermodynamic limit.
The inset of (c) presents the system size scaling of the two-point correlator at various $\theta$, as given by Eq.~\eqref{eq:2pt}.
}
\label{fig:2pt}
\end{figure}

\emph{Computation scheme.--}The fuzzy sphere provides a realization of 3D CFT on the geometry $S^2\times \mathbb R$, allowing us to utilize the state-operator correspondence \cite{Cardy1984,Cardy1985} to simplify the computation of the four-point correlator.
We begin by choosing the conventional conformal frame for the four-point correlator in Eq.~\eqref{eq:4pt_definition}, where we set $\bm x_1=(0, 0, 0)$, $\bm x_2=(x, y, 0)$, $\bm x_3=(1, 0, 0)$, and $\bm x_4 = (\infty, 0, 0)$.
We can introduce the complex coordinate $z=x+iy$, $\bar z=x-iy$, so that the crossing ratios are $u=z\bar z$ and $v=(1-z)(1-\bar z)$.
Next, we employ the state-operator correspondence to convert operators to states \cite{yellowbook}, i.e., $\phi_1(\bm x_1=0)|0\rangle = |\phi_1\rangle$ and $\lim_{\bm x_4\rightarrow\infty} x_4^{2\Delta} \langle 0| \phi_4(\bm x_4) = \langle\phi_4|$.
As a result, the computation of Eq. \eqref{eq:4pt_definition} reduces to evaluating a ``two-operator'' expectation value $\langle \phi_4 |\phi_2 (\bm x_2) \phi_3 (\bm x_3) |\phi_1 \rangle$.
It is more convenient to parameterize the coordinate $(z=re^{i\theta}, \bar z=re^{-i\theta})$ by spherical coordinates, and we will use $(z, \bar z)$ and $(r, \theta)$ interchangeably.
As shown in Fig.~\ref{fig:2pt}(a), in spherical coordinates, we can position $\bm x_3$ at the north pole of the unit sphere $(r=1, \theta=0, \varphi=0)$ and $\bm x_2$ at the point $(r, \theta, \varphi=0)$.
Moreover, since we are performing computations in the geometry $S^2\times \mathbb R$, an additional Weyl transformation $(r, \theta,\varphi) \rightarrow (\tau = R \log r, \theta, \varphi)$ is required to map our results to correlators in Euclidean $\mathbb R^3$ ($R$ is the radius of the sphere $S^2$ on the cylinder).
As explained in Supplementary Material Sec. B, the four-point correlator can be expressed as the $S^2\times \mathbb R$ observable:
\begin{equation}\label{eq:4pt}
g_{\phi_1\phi_2\phi_3\phi_4}(r,\theta) = \frac{r^{\Delta_1} \langle \phi_4 | \phi_2(\tau=R\log r, \theta) \phi_3 |\phi_1\rangle}{\langle 0| \phi_2 |\phi_2\rangle \langle 0| \phi_3 |\phi_3\rangle}.
\end{equation}
Here, and in subsequent expressions, we omit writing a coordinate ($\tau$, $\theta$, or $\varphi$) explicitly when it is taken to be zero.

In the fuzzy sphere model, the eigenstates of the Hamiltonian correspond to CFT states and are nearly exact, with small finite-size corrections. On the other hand, we can construct local operators to approximate CFT primary operators and compute their correlators.
For instance, to approximate the primary operator $\sigma$, we can use the $\mathbb{Z}_2$ odd density operator $n^z(\tau, \theta, \varphi)$ as an approximation \cite{hu2023operator}. The correlator can then be computed as follows:
\begin{equation}\label{eq:4pt_compute}
g_{\phi_1\sigma\sigma\phi_4}(r, \theta) = \frac{r^{\Delta_1} \langle \phi_4 | n^z(\tau = R \log r, \theta) n^z |\phi_1\rangle}{\langle 0 | n^z |\sigma\rangle^2} + \mathcal{O}(R^{-1}),
\end{equation}
where the subleading contribution comes from the component of the descendant operator $\partial_\mu \sigma$ in $n^z$~\cite{hu2023operator}.
This subleading contribution becomes negligible in the limit of a large system size ($R \gg 1$).

In practice, all the computations are performed in the orbital space defined by the fermionic operator $c_{m,\uparrow(\downarrow)}$ with $m=-s, -s+1, \cdots, s$, where $ s$ is monopole charge at the origin of the sphere \cite{Sphere_LL_Haldane}, and these fermion operators form a spin-$s$ irreducible representation of the $SO(3)$ sphere rotation. 
We can easily translate between real space and orbital space by using monopole harmonics $Y^{(Q)}_{s,m}$ \cite{WuYangmonopole}
\begin{equation}\label{eq:LLLprojection}
\psi_a (\Omega) = \sum_{m=-s}^s c_{m, a} \bar Y^{(s)}_{s, m} (\Omega),
\end{equation}
and subsequently take the limit $s\rightarrow\infty$ to approach the thermodynamic limit.
So the monopole flux $s$ or equivalently the electron number $N=2s+1$ plays the role of system size (i.e. space volume $\sim R^2$).
In this context, the density operator becomes
\begin{align}
n^a(\theta,\varphi) 
&= \sum_{m_1,m_2} Y^{(s)}_{s,m_1}(\theta,\varphi) \bar Y^{(s)}_{s,m_1}(\theta,\varphi) \bm c_{m_1}^\dag \sigma^a \bm c_{m_2} \nonumber \\
&= \sum_{l=0}^{2 s} \sum_{m=-l}^l n_{l, m}^a Y_{l, m}(\theta,\varphi),
\end{align}
where $ Y_{l, m}$ is the spherical harmonics relating to monopole harmonics $ Y^{(Q=0)}_{l, m}$.

The numerator and denominator of Eq.~\eqref{eq:4pt_compute} can be computed separately by 
\begin{align}
\langle  0| n^z |\sigma\rangle = \sum_{l,m} Y_{l,m}(\theta&,\varphi) \langle  0| n^z_{l,m}|\sigma\rangle = \frac{1}{\sqrt{4\pi}} \langle 0 | n^z_{l=m=0} |\sigma \rangle \nonumber \\
\langle \phi_4| n^z(\tau, \theta) n^z |\phi_1 \rangle=&\sum_{l=0}^{2s} \bar{Y}_{l,m=0}(\theta, 0)Y_{l,m=0}(0, 0) \nonumber \\
&\langle \phi_4|n^z_{l,m=0}(\tau) n^z_{l,m=0}|\phi_1\rangle.
\end{align}
Note that the spherical harmonics  $Y_{l,m}(0,0)$ is non-zero only for $m=0$ which greatly simplifies the calculation. 
The major computation is to evaluate 
\begin{equation}\label{eq:ite}
\langle \phi_4|n^z_{l,m=0}(\tau) n^z_{l,m=0}|\phi_1\rangle = \langle\phi_4|e^{\tau H}n^z_{l,m=0}e^{-\tau H} n^z_{l,m=0}|\phi_1\rangle.    
\end{equation}
In this work, we use the matrix product state (i.e., density-matrix renormalization group) to perform the numerical simulations~\cite{SWhite1992,Feiguin2008}, and for the imaginary time evolution, we employ the time-dependent variational principle (TDVP) algorithm \cite{TDVP2011, itensor}. The numerical errors are controlled by the matrix product state bond dimension $D$ and the time evolution step $\Delta \tau$ (i.e., $\Delta r$).
We have compared different values of $\Delta \tau$ (i.e., $\Delta r$) and $D$, which yield consistent results. For the time-evolution simulation, we choose $D = 3000$ and evolve from $r = 1$ to $r = 4$ with 100 steps, i.e., $\Delta r = 0.03$.
For the static simulation, specifically when $r = 1$ or equivalently $\tau = 0$, we can achieve a larger system size by using a larger value of $D = 7000$.

As one can observe, the computed correlators are continuous functions of $\theta$, demonstrating that the fuzzy sphere model is defined in the continuum. We obtain a series expansion of $\cos\theta$, given by $g_{\phi_1\phi_2\phi_3\phi_4} (r,\theta) = \sum_{n=0}^{2s} a_n(r) \cos^n\theta$, where $a_n(r)$ is a numerical factor evaluated numerically. The series expansion is truncated at the order of $2s$, implying that the singularity at $(r=1, \theta=0)$ will only be fully recovered in the limit $s\rightarrow\infty$. Importantly, the numerical factor $a_n(r)$ is also a continuous function of $r$. In practice, we have to target at specific values of $r$, although our calculations allow for arbitrary values of $r$ to be accessed.

Before presenting our numerical results, it is important to provide a precise definition of the sphere radius $R$ for the computation of time-evolution (i.e. $\tau=R\log r$) in Eq.~\eqref{eq:4pt_compute}. 
$R$ can be defined by relating the Hamiltonian $H$ (after shifting the groundstate energy to 0) with the CFT dilatation operator $\hat D$, $H= \hat D/R$.  Consequently, for a system of a specific size $N=2s+1$, we define the sphere radius $R$ as $R=\Delta_\phi/\delta E_\phi$, where $\phi$ represents a CFT operator (which can be chosen as either the primary operator $\sigma$ or the stress tensor), and $\delta E_\phi$ denotes the energy gap of the excited state associated with $\phi$ in the context of the state-operator correspondence.
We also note that, except for the computation of time-evolution, we will interchangeably use $R$ and $\sqrt{N}$, and perform the finite size extrapolation using $\sqrt{N}$.

\emph{Tow-point correlator.--}In Eq. \eqref{eq:4pt}, if we choose the states $|\phi_1\rangle$ and $|\phi_4\rangle$ to be the ground state, we obtain the expression for the two-point correlator, which is known exactly as (see Supplementary Material Section B):
\begin{equation}\label{eq:2pt}
g_{\phi \phi}(r,\theta) = \frac{ r^\Delta}{(r^2+1-2r\cos\theta)^\Delta}.
\end{equation}
From a computational perspective (as seen in Eq. \eqref{eq:4pt_compute}), there is no fundamental difference between the computation of the two-point and four-point correlators. Therefore, we will begin by computing the two-point correlator as a numerical benchmark for subsequent calculations of the four-point correlator.

The two-point correlator given by Eq. \eqref{eq:2pt} can also be utilized to accurately determine the critical point of the system.
In particular, the equal-time correlator $g_{\sigma\sigma}(r=1, \theta) = (2\sin(\theta/2))^{-2\Delta_\sigma}$ is a dimensionless function that solely depends on the angle $\theta$ between the two operators.
Thus, the critical point can be identified by finding the value of $h$ that yields the best fit of this dimensionless function. Specifically, we set $\theta=\pi$, resulting in $g_{\sigma\sigma}(r=1, \theta=\pi) = 2^{-2\Delta_\sigma}\approx 0.487577$ ($\Delta_\sigma\approx 0.5181489(10)$~\cite{Kos:2016ysd}). We can then examine which value of $h$ produces the correct $g_{\sigma\sigma}(r=1, \theta=\pi)$ in the thermodynamic limit.
Fig.\ref{fig:2pt}(b) displays $g_{\sigma\sigma}(r=1, \theta=\pi)$ for different values of $h$ (specifically, $h=3.155, 3.16, 3.165$). By employing proper finite-size extrapolation with respect to $1/\sqrt{N}$, we find that $h_c\approx 3.16$.  This result is consistent with the previously determined critical point using local order parameter scaling~\cite{ZHHHH2022}.
Below we show results of correlators at $h=3.16$.

Fig.~\ref{fig:2pt}(c-d) depicts the two-point correlation function $g_{\sigma\sigma}(r,\theta)$ by fixing $r=1$ and $r=1.6$, respectively, for different system sizes $N=16-40$, and $N=16-32$.
The $\theta$-dependence of two-point correlator is quite close to the CFT prediction Eq.~\eqref{eq:2pt}, except for the small $\theta$ regime where singular behavior is anticipated and can only be produced at infinite $s$ limit.

\begin{figure}[t]
\includegraphics[width=0.32\linewidth]{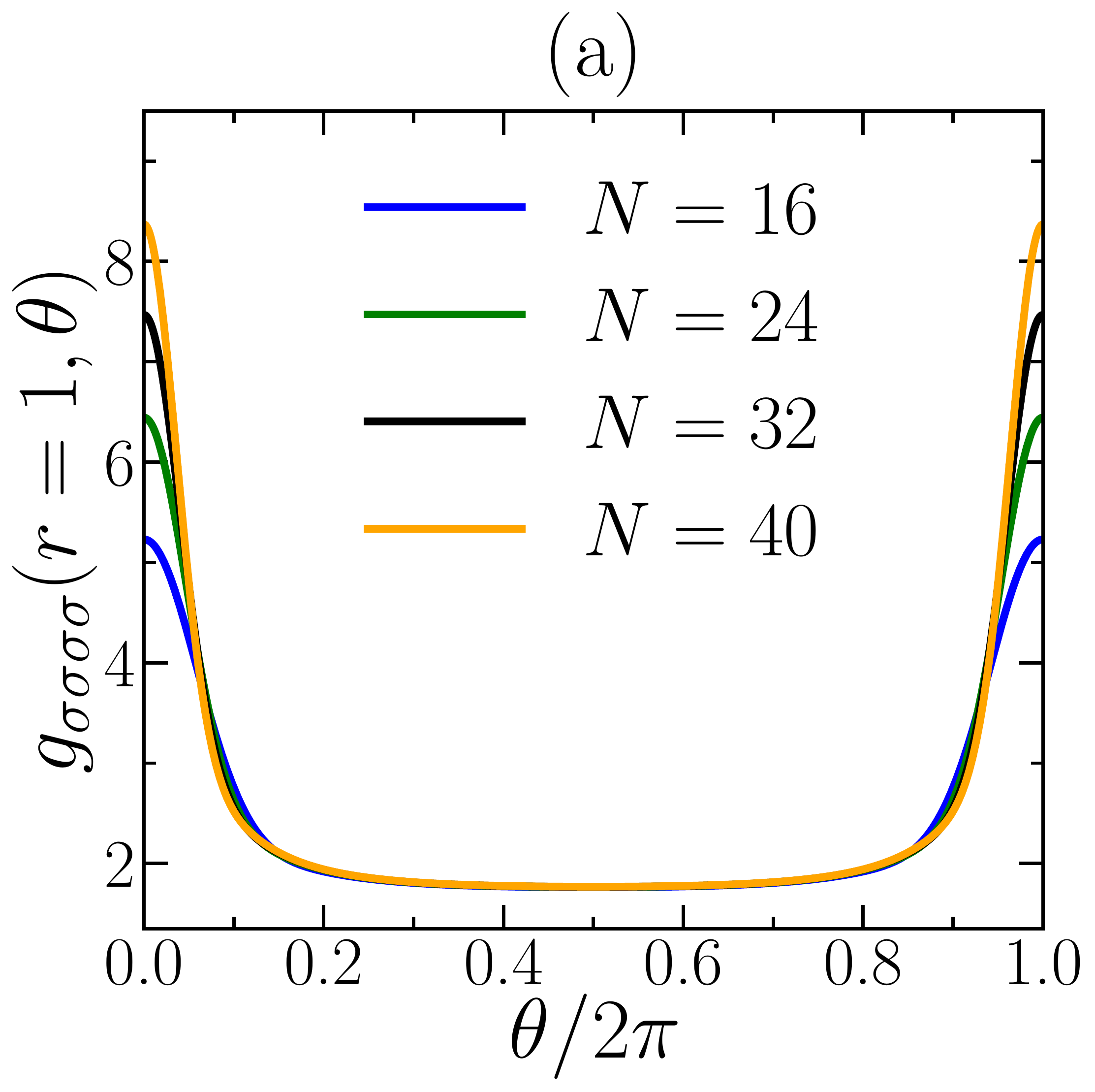}
\includegraphics[width=0.32\linewidth]{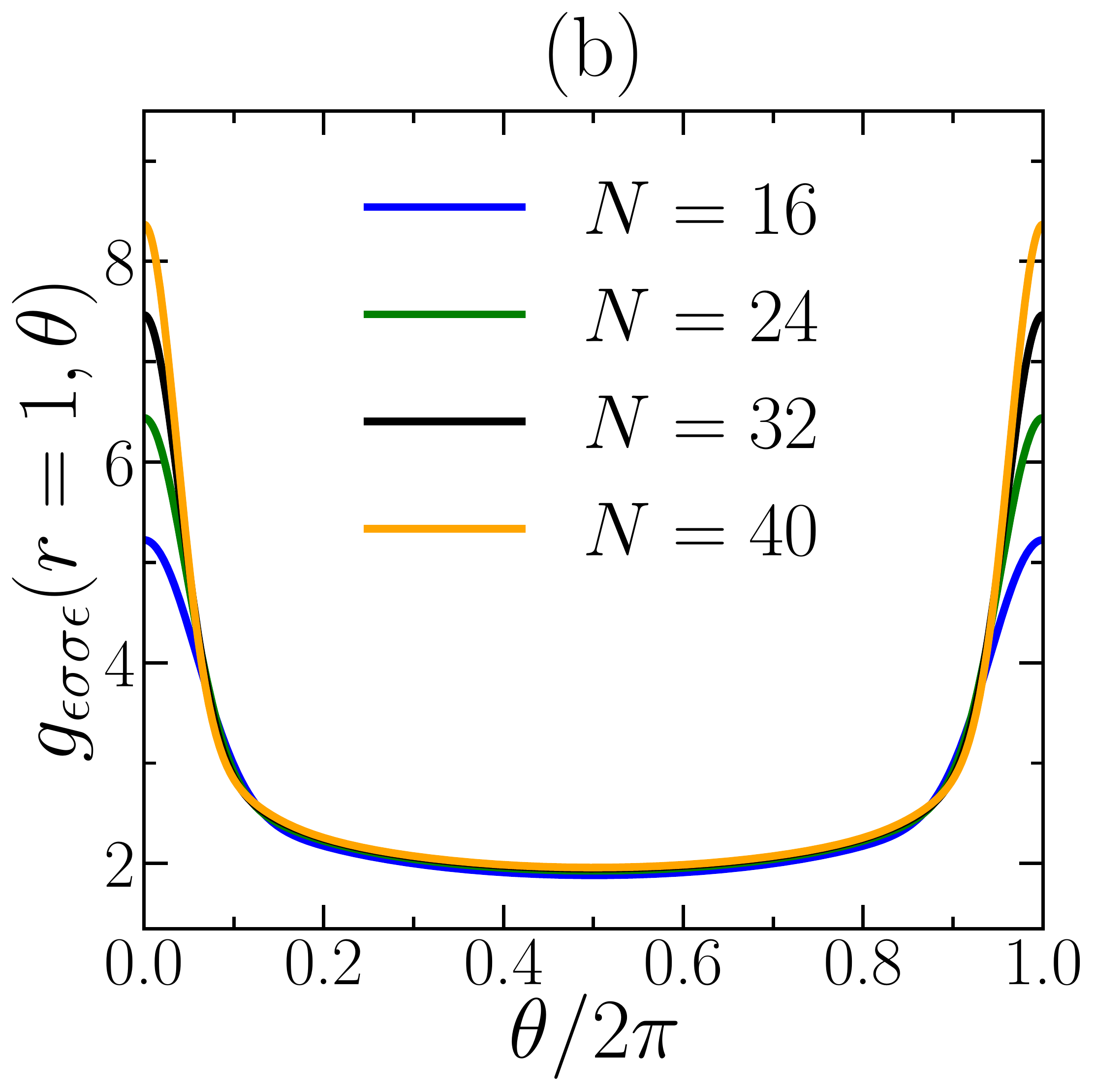}
\includegraphics[width=0.32\linewidth]{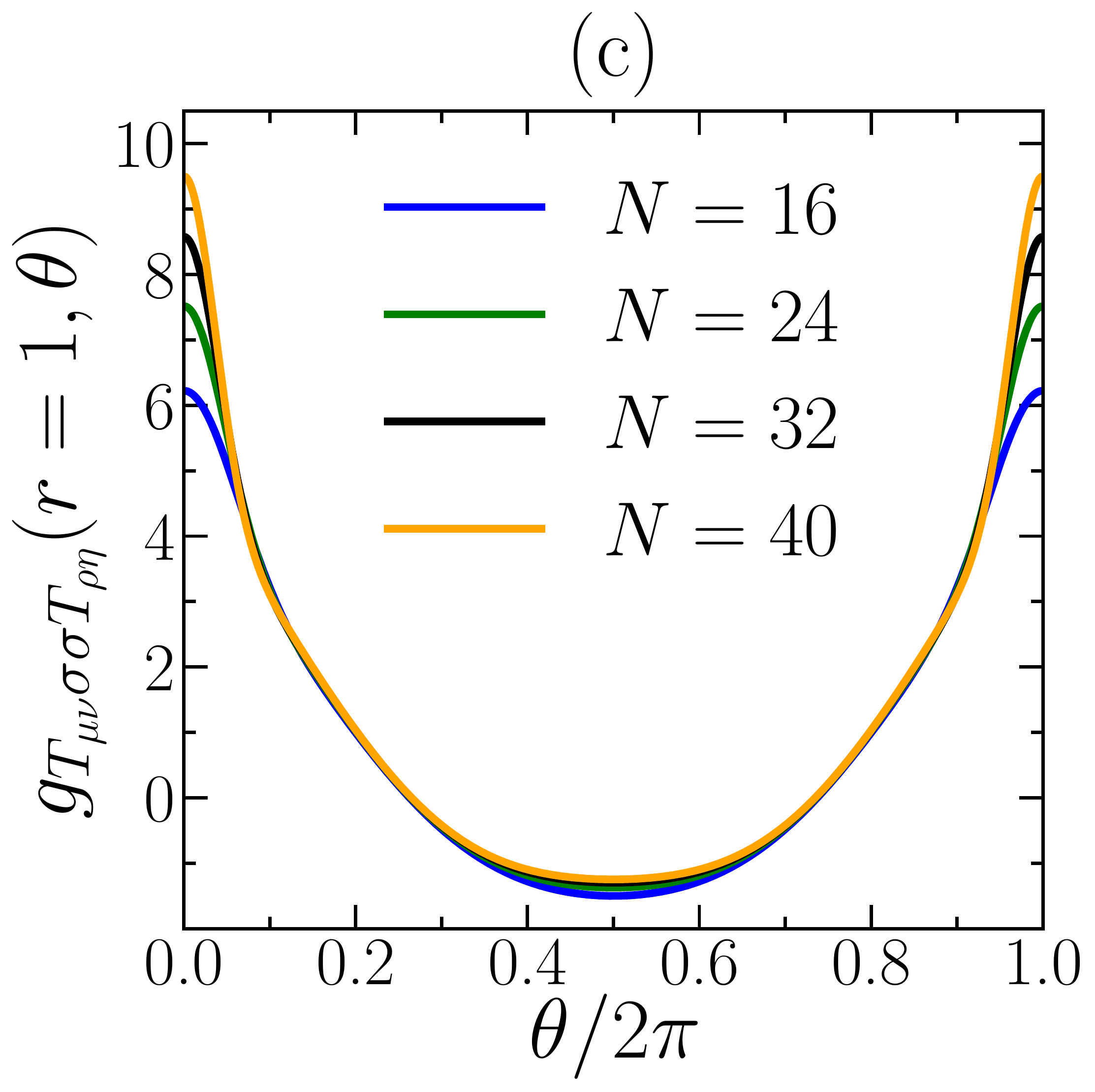}
\caption{Angle dependence of the four-point correlator at $r=1$ is shown for (a) $g_{\sigma\sigma\sigma\sigma}(r=1, \theta)$, (b) $g_{\epsilon\sigma\sigma\epsilon}(r=1, \theta)$, and (c) $g_{T_{\mu\nu}\sigma\sigma T_{\rho\eta}}(r=1, \theta)$. The curves represent different system sizes ranging from $N=16$ to $N=40$, and they converge to each other. The discrepancy between the curves becomes more significant for smaller values of $\theta$, which is attributed to the singularity at $\theta=0$ in the thermodynamic limit.
}
\label{fig:Static_4pt}
\end{figure}

\begin{table}[b]
\caption{\label{table:4pt}  Comparison of $g_{\sigma\sigma\sigma\sigma}(r=1,\theta)$ between our fuzzy sphere computation and the conformal bootstrap data reveals very small discrepancies. The conformal bootstrap data is obtained by reconstructing the four-point correlator using a conformal block expansion, for example, see Ref.~\cite{Rychkov2017_4pt}. These discrepancies diminish as the system size $N=2s+1$ increases, indicating good agreement between our computation and the conformal bootstrap results.}
\setlength{\tabcolsep}{0.2cm}
\renewcommand{\arraystretch}{1.4}
\centering
\begin{tabular}{cccccc} 
\hline\hline & Bootstrap & $N=40$ & $N=32$ & $N=24$ & $N=16$ \\ 
$\theta=\pi$ & 1.76855 & 1.76742 & 1.76671 & 1.76549 & 1.76244 \\
$\theta=\pi/3$ & 2.049 & 2.03921 & 2.03495 & 2.02470 & 2.01212 \\
\hline\hline
\end{tabular} 
\end{table}

\begin{figure}[t]
\includegraphics[width=0.49\linewidth]
{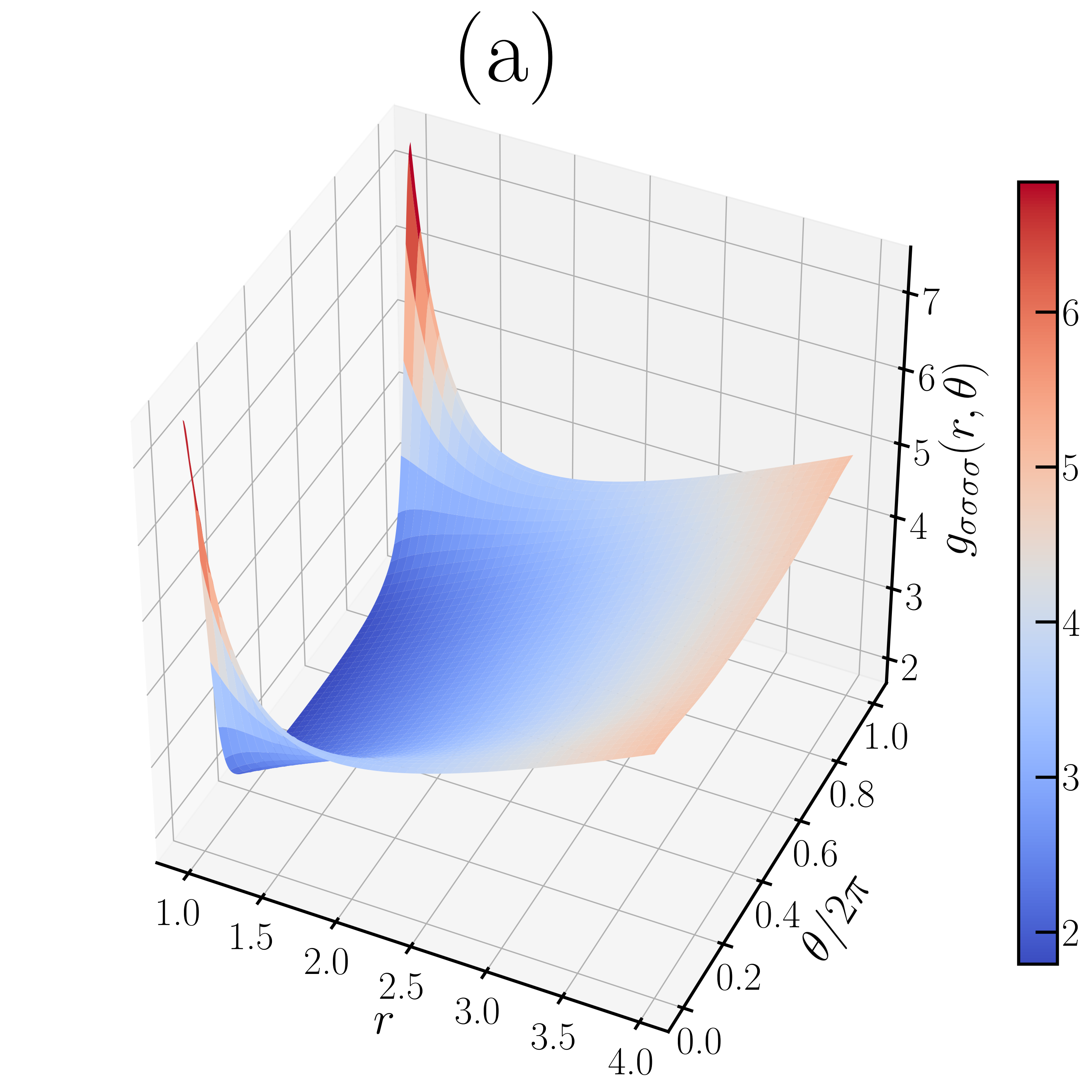}
\includegraphics[width=0.49\linewidth]{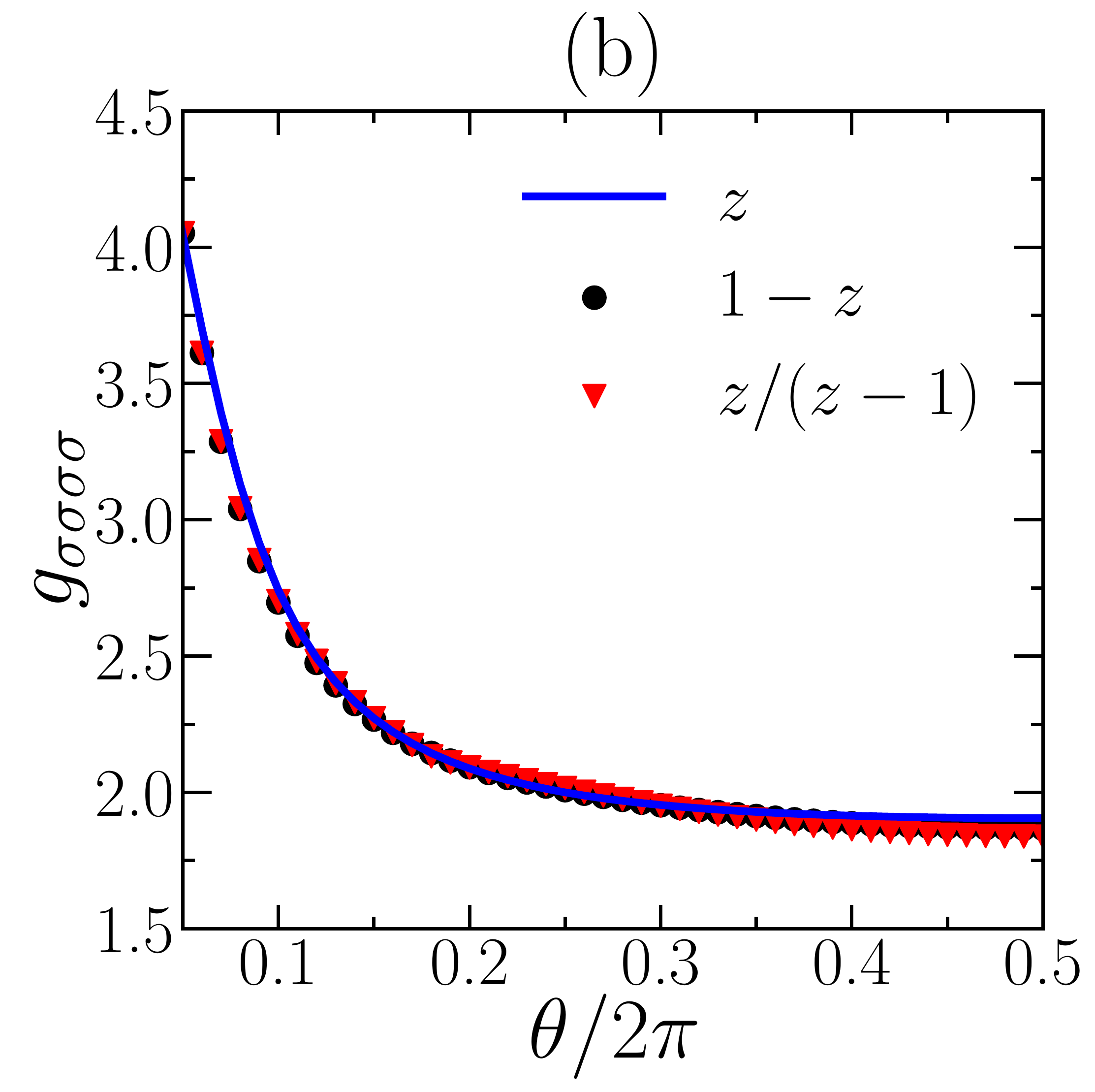}
\caption{(a) A 3D plot of the four-point correlator $g_{\sigma\sigma\sigma\sigma}(r, \theta)$ is shown for a system size of $N=32$.
(b) The crossing symmetry of the four-point correlator $g_{\sigma\sigma\sigma\sigma}$ is demonstrated for a system size of $N=32$. The blue solid line represents $g(z,\bar z)$ obtained by setting $z=re^{i\theta}$ with $r=1.15$. The quantities $\frac{|z|^{2\Delta}}{|1-z|^{2\Delta}}g(1-z,1-\bar z)$ and $g\left(\frac{z}{z-1},\frac{\bar z}{\bar z-1}\right)$ are plotted as black circles and red triangles, respectively. These data points agree well with $g(z, \bar z)$, providing support for the crossing symmetry expressed in Eqs.\eqref{eq:cs1} and\eqref{eq:cs2}.
}
\label{fig:cs}
\end{figure}

\emph{Four-point correlator.--}Next, we turn to the computation of the four-point correlator $g_{\phi_1\sigma\sigma\phi_4}(r,\theta)$, which can be achieved by selecting the excited states corresponding to the desired CFT operators as $|\phi_1\rangle$ and $|\phi_4\rangle$ in Eq. \eqref{eq:4pt_compute}.
Fig.\ref{fig:Static_4pt}(a-c) presents the results for $g_{\sigma \sigma \sigma \sigma} (r=1,\theta)$, $g_{\epsilon \sigma \sigma \epsilon} (r=1,\theta)$, and $g_{T_{\mu\nu} \sigma \sigma T_{\rho\eta}} (r=1,\theta)~$\footnote{Here $g_{T_{\mu\nu} \sigma \sigma T_{\rho\eta}} (r=1,\theta)$ is evaluated in a specific tensor structure where we choose the two CFT states to be $|T_{\mu\nu}, m=0\rangle= \lim_{r\rightarrow 0} \sqrt{\frac{2}{27}} (2T_{33}(\bm r) - T_{11}(\bm r)-T_{22}(\bm r)) |0\rangle$.}, and Fig.~\ref{fig:cs}(a) shows $g_{\sigma\sigma\sigma\sigma}(r,\theta)$ for the regime $1\le r\le 4$. It is evident that all the curves of different system sizes $N=16-40$ in Fig.\ref{fig:Static_4pt} converge quickly in the intermediate $\theta$ regime ($\theta\in [0.4\pi, 1.6\pi]$), and the small $\theta$ results are improved at a large system size.
Furthermore, we compare our results at $\theta=\pi, \pi/3$ with the four-point correlator obtained from conformal bootstrap data in Table~\ref{table:4pt}, demonstrating excellent agreement between them.

An important feature of the four-point correlator is the crossing symmetry. 
For identical scalar primaries (e.g., $\sigma$) there are two independent crossing equations,
\begin{equation}\label{eq:cs1}
\frac{g(z, \bar z)}{(z\bar z)^{\Delta}} = \frac{g(1-z, 1-\bar z)}{((1-z)(1-\bar z))^\Delta},
\end{equation}
and 
\begin{equation}\label{eq:cs2}
g(z, \bar z) = g\left(\frac{z}{z-1},\frac{\bar z}{\bar z-1}\right).
\end{equation}
Our data for $g(z,\bar z)$, $\frac{|z|^{2\Delta}}{|1-z|^{2\Delta}}g(1-z,1-\bar z)$, and $g\left(\frac{z}{z-1},\frac{\bar z}{\bar z-1}\right)$ in Fig.~\ref{fig:cs}(b) exhibit excellent agreement among the three, providing strong evidence for the crossing symmetry of our computed four-point correlator.

\emph{Summary and discussion.--}Using the fuzzy sphere regularization, we have introduced a novel approach to study 3D CFT correlators. A key feature of this method is that it produces correlators that are continuous functions of spacetime coordinates. 
More concretely, we introduced a dimensionless two-point correlator, analogous to the well-known binder cumulant, which can serve as a valuable tool for precisely determining critical points in various phase transitions within the fuzzy sphere model.
We further directly compute several four-point correlators ($\langle\sigma\sigma\sigma\sigma\rangle$, $\langle \sigma \sigma \epsilon\epsilon\rangle$, and $\langle \sigma\sigma T_{\mu\nu} T_{\rho\eta}\rangle$) of the paradigmatic 3D Ising CFT, and verified the crossing symmetry of $\langle \sigma\sigma\sigma\sigma\rangle$.
Our results for the four-point correlator $\langle \sigma\sigma\sigma\sigma\rangle$ exhibit excellent agreement with indirect reconstruction using conformal bootstrap data. For instance, at $z=\bar z=-1$, the discrepancy is approximately $0.06\%$ for $N=40$ electrons (spins)!

One intriguing future direction is to directly extract conformal data, such as primary operator scaling dimensions and operator product expansion coefficients, using the inversion formula~\cite{Simon2017Inversion}.
It is also interesting to compute multi-point correlator such as 5-point~\cite{Poland2023_5point} and 6-point correlator, in particular, the latter provides access to all the 3-point structures.
Another exciting avenue to pursue is to study 3D CFT correlators in Lorentzian spacetime. This involves studying quantities such as the out-of-time correlator, which captures the real-time dynamics and information scrambling properties of the CFT.

\textit{Acknowlegement.---}
We thank Shahnewaz Ahmed, Davide Gaiotto, Ning Su, Rokas Veitas, Zheng Zhou for helpful discussions. This work was supported by National Science Foundation of China under No. 92165102, 11974288 (L.D.H.,W.Z.), and by the R$\&$D Program of Zhejiang under No. 2022SDXHDX0005 (C.H.). Research at Perimeter Institute is supported in part by the Government of Canada through the Department of Innovation, Science and Industry Canada and by the Province of Ontario through the Ministry of Colleges and Universities.
Y.C.H. thanks KITP for hospitality where part of this work was completed. This research was supported in part by the National Science Foundation under Grant No. NSF PHY-1748958.

\bibliography{ref}

\clearpage

\appendix

\begin{widetext}

\begin{center}
\textbf{Supplementary material}    
\end{center}

In this supplementary material, we will show more details about the correlators of CFT scalar primary operators on the cylinder geometry $S^{2}\times \mathbb{R}$. For ease of notation, we will explicitly consider the case of three dimensions, although most derivations directly apply to other dimensions. We will use the Weyl transformation \cite{Cardy1984,Cardy1985},
\begin{equation}
(r, \theta, \varphi) \rightarrow (\tau = R\log r, \theta, \varphi),
\end{equation}
to map the correlators in Euclidean space $\mathbb{R}^3$ to the correlators on the cylinder $S^{2}\times \mathbb{R}$. Here, $(\theta, \varphi)$ are the spherical angles on $S^2$, and $R$ is the radius of the sphere $S^2$ on the cylinder.
The operator $\phi_{flat}(r, \theta, \varphi)$ in $\mathbb{R}^3$ and the operator $\phi_{cyl}(\tau, \theta, \varphi)$ in $S^2\times \mathbb{R}$ are related by the equation,
\begin{equation}
\phi_{cyl}(\tau,\theta, \varphi) = R^{-\Delta} e^{\tau\Delta/R} \phi_{flat}(r, \theta, \varphi),
\end{equation}
where $\Delta$ is the scaling dimension of the operator.

\section{A. Two-point correlator $g_{\phi \phi}$ }
Let us start with the two-point correlator. In $\mathbb R^3$, the two-point correlator of a scalar primary is given by
\begin{equation} \label{eq:2ptflat}
\left\langle \phi_{flat}\left(r, \theta, \varphi\right) \phi_{flat}\right\rangle=\frac{1}{\left(r^2+1-2 r \cos \theta\right)^{\Delta}}.
\end{equation}
where we have considered $r_2$ on the unit sphere, i.e., $r_2=1$. After performing the Weyl transformation, the resulting two-point correlator on the cylinder $S^{2}\times \mathbb{R}$ is given by
\begin{equation} \label{eq:2ptcylinder}
\langle \phi_{cyl}(\tau=R\log r, \theta, \varphi) \phi_{cyl} \rangle = \frac{R^{-2\Delta} r^\Delta}{(r^2+1-2r\cos\theta)^\Delta}.
\end{equation}

We can also use the state-operator correspondence
\begin{equation} \label{eq:stateop}
|\phi\rangle=\lim_{r \rightarrow 0} {\phi}(r, \theta,\varphi)|0\rangle, \quad\langle\phi|=\lim_{r \rightarrow \infty} r^{2 \Delta}\langle 0| {\phi}(r, \theta,\varphi),
\end{equation}
to map the operator $\phi_{flat}(r_2)$ in Eq.~\eqref{eq:2ptflat} to a state $|\phi\rangle$ by taking $r_2\rightarrow 0$,
\begin{equation}
\langle 0| {\phi}_{flat}\left(r, \theta, \varphi\right) |\phi\rangle = \frac{1}{r^{2\Delta}},
\end{equation}
and the Weyl transformation gives
\begin{equation} \label{eq:op_state}
\langle  0| {\phi}_{cyl}\left(\tau=R\log r, \theta, \varphi\right) |\phi\rangle = R^{-\Delta} r^{-\Delta},
\end{equation}
which is independent of the spherical angles. Combining Eq.~\eqref{eq:2ptcylinder} and \eqref{eq:op_state}, we obtain a  two-point correlator:
\begin{equation}\label{eq:twopoint_final}
g_{\phi\phi}(r,\theta)=\frac{\langle \phi_{cyl}(\tau=R\log r, \theta, \varphi) \phi_{cyl} \rangle}{\langle  0| {\phi}_{cyl} |\phi\rangle^2} = \frac{ r^\Delta}{(r^2+1-2r\cos\theta)^\Delta}.
\end{equation}
This form of the two-point correlator on the sphere is totally fixed by the conformal symmetry. For convenience, we skip writing the subscripts "cyl" and "flat" in the following and the main text.

\section{B. Four-point correlator $g_{\phi_1 \phi_2 \phi_3 \phi_4}$ }
Next we turn to the four-point correlator. The general four-point correlator of scalar primaries in Euclidean space $\mathbb{R}^3$ has the following functional form
\begin{equation}
\begin{aligned}
\langle {\phi_4}(\bm x_4) {\phi_3}(\bm x_3) {\phi_2}(\bm x_2) {\phi_1}(\bm x_1)\rangle=&\left(\frac{x_{13}}{x_{14}}\right)^{\Delta_4-\Delta_3}\left(\frac{x_{14}}{x_{24}}\right)^{\Delta_2-\Delta_1} \\
&\frac{g_{\phi_1 \phi_2 \phi_3 \phi_4}(u, v)}{x_{34}^{\Delta_4+\Delta_3} x_{12}^{\Delta_2+\Delta_1}}, 
\end{aligned}
\end{equation}
where $u=\frac{x_{12}^2 x_{34}^2}{x_{13}^2 x_{24}^2}$ and $v=\frac{x_{14}^2 x_{23}^2}{x_{13}^2 x_{24}^2}$ are called crossing ratios. After choosing the conventional conformal frame by setting $\bm x_1=(0, 0, 0)$, $\bm x_2=(x, y, 0)$, $\bm x_3=(1, 0, 0)$, and $\bm x_4 = (\infty, 0, 0)$, we obtain
\begin{equation}
g_{\phi_1 \phi_2 \phi_3 \phi_4}(u, v)=r^{\Delta_2+\Delta_1}\langle {\phi_4}| {\phi_2}(\bm x_2) {\phi_3}(\bm x_3) |{\phi_1}\rangle,
\end{equation}
where the state-operator correspondence $\phi_1(\bm x_1=0)|0\rangle = |\phi_1\rangle$ and $\lim_{\bm x_4\rightarrow\infty} x_4^{2\Delta} \langle 0| \phi_4(\bm x_4) = \langle\phi_4|$ has been used. Similar to the two-point correlator, we use the Weyl transformation to get the final four-point correlator on the cylinder $S^{2}\times \mathbb{R}$
\begin{equation}
g_{\phi_1\phi_2\phi_3\phi_4}(r,\theta) = \frac{r^{\Delta_1} \langle \phi_4 | \phi_2(\tau=R\log r, \theta) \phi_3 |\phi_1\rangle}{\langle 0| \phi_2 |\phi_2\rangle \langle 0| \phi_3 |\phi_3\rangle}.
\end{equation}

\section{C. Crossing symmetry of four-point correlator }
For the sake of convenience, we will explicitly consider identical scalar primaries, i.e., $\Delta_i=\Delta, \phi_i=\phi, i=1,2,3,4$. Therefore, we have a simpler form of the four-point correlator
\begin{equation}
\langle {\phi}(\bm x_4) {\phi}(\bm x_3) {\phi}(\bm x_2) {\phi}(\bm x_1) \rangle = \frac{g(u, v)}{x_{12}^{2\Delta} x_{34}^{2\Delta} }.
\end{equation}
The ordering of fields within correlators does not matter, therefore, we can freely interchange them. For example, if we exchange $\bm x_1\leftrightarrow \bm x_3$, we can obtain $\tilde{u}=v, \tilde{v}=u$ and, 
\begin{equation}
\frac{g(u, v)}{x_{12}^{2\Delta} x_{34}^{2\Delta} }= \frac{g(\tilde{u}, \tilde{v})}{x_{23}^{2\Delta} x_{14}^{2\Delta} },
\end{equation}
i.e., 
\begin{equation}
u^{-\Delta} g(u, v) = v^{-\Delta} g(v, u).
\end{equation}
We could also exchange $\bm x_1\leftrightarrow \bm x_2$ and get 
$\tilde{u}=\frac{x_{12}^2 x_{34}^2}{x_{14}^2 x_{23}^2}=\frac{u}{v}, \tilde{v}=\frac{x_{13}^2 x_{24}^2}{x_{14}^2 x_{23}^2}=\frac{1}{v}$. Thus, we obtain the second crossing equation
\begin{equation}
\frac{g(u, v)}{x_{12}^{2\Delta} x_{34}^{2\Delta} }= \frac{g(\tilde{u}, \tilde{v})}{x_{12}^{2\Delta} x_{34}^{2\Delta} },
\end{equation}
i.e., 
\begin{equation}
g(u, v) = g\left(\frac{u}{v}, \frac{1}{v}\right).
\end{equation}
Similarly, the third crossing equation comes from the exchange $\bm x_1\leftrightarrow \bm x_4$. We get $\tilde{u}=\frac{x_{24}^2 x_{13}^2}{x_{34}^2 x_{12}^2}=\frac{1}{u}, \tilde{v}=\frac{x_{14}^2 x_{23}^2}{x_{34}^2 x_{12}^2}=\frac{v}{u}$, and 
\begin{equation}
\frac{g(u, v)}{x_{12}^{2\Delta} x_{34}^{2\Delta} }= \frac{g(\tilde{u}, \tilde{v})}{x_{24}^{2\Delta} x_{13}^{2\Delta} },
\end{equation}
i.e.,
\begin{equation}
g(u, v) = u^\Delta g\left(\frac{1}{u}, \frac{v}{u}\right).
\end{equation}
In terms of the conformal frame, $\bm x_1=(0, 0, 0)$, $\bm x_2=(x, y, 0)$, $\bm x_3=(1, 0, 0)$, and $\bm x_4 = (\infty, 0, 0)$, where $z=x+iy=re^{i\theta}$,  $\bar z=x-iy=re^{-i\theta}$, the crossing symmetry becomes
\begin{equation}
\frac{g(z, \bar z)}{(z\bar z)^{\Delta}} = \frac{g(1-z, 1-\bar z)}{((1-z)(1-\bar z))^\Delta},
\end{equation}
and 
\begin{equation}
g(z, \bar z) = g\left(\frac{z}{z-1},\frac{\bar z}{\bar z-1}\right),
\end{equation}
and 
\begin{equation}
g(z, \bar z) = (z\bar z)^{\Delta}g\left(\frac{1}{z},\frac{1}{\bar z}\right).
\end{equation}

\end{widetext}

\end{document}